\documentclass[aps,prl,twocolumn,superscriptaddress]{revtex4}
\usepackage{graphicx}
\usepackage{amsmath,amssymb}
\begin{document}

\title{Subwavelength optical spatial solitons and three-dimensional localization in disordered ferroelectrics: towards metamaterials of nonlinear origin}

\author{C.Conti} 
\affiliation{Dep. Molecular Medicine, Univ. Sapienza, Viale Regina Elena, 324, Univ. Sapienza, 00161 Rome (IT)}
\affiliation{Institute for Complex Systems - CNR, Dep. Physics University Sapienza,  Piazzale Aldo Moro 2, 00185 Rome, Italy}
\author{A. J. Agranat} 
\affiliation{ Applied Physics Department, Hebrew University of Jerusalem, 91904, Israel}
\author{E. DelRe} 
\affiliation{Dipartimento  di Ingegneria Elettrica e
  dell'Informazione, Universita' dell'Aquila, 67100 L'Aquila, Italy}
\date{\today}

\begin{abstract}
\noindent We predict the existence of a novel class of multidimensional
light localizations in out-of-equilibrium ferroelectric crystals. 
In two dimensions, the non-diffracting beams form at arbitrary low power level and propagate even when their width is well below the optical wavelength. 
In three dimensions, a novel form of subwavelength light bullets is found.  The effects emerge when compositionally disordered crystals are brought to their metastable glassy state, and can have a profound impact on super-resolved imaging and ultra-dense optical storage, while resembling many features of 
the so-called metamaterials, as the suppression of evanescent waves. 
\end{abstract}
\maketitle

{\it Introduction ---}
Out-of-equilibrium materials display remarkable features, most of them still to be understood. Recent experiments in supercooled photorefractive crystals have allowed the observation of ``scale-free'' optical solitons \cite{DelRe11} supported by an extremely weak diffusive nonlinearity \cite{Crosignani99,DelRe09, DelRe09book}, which becomes active through the emergence of a dipolar glass with anomalously enhanced susceptibility. These nonlinear beams have a truly remarkable feature: they are independent of size and intensity.   
Size-independence spurns a very basic and potentially ground-breaking exploration: are these scale-free optical beams capable of propagating even when their size is noticeably smaller than the optical wavelength?   

In this Letter we predict that glassy photorefractive ferroelectrics \cite{Agranat92, Samara03, Bokov06} support 
a novel kind of multidimensional light localization at scales below the optical wavelength. 
The effect requires a huge susceptibility that can be harnessed in the out-of-equilibrium, or non-ergodic, phase, by acting on the previous history of the sample \cite{DelRe11}. The finding is thus part of a still infant field of investigation that focuses on using out-of-equilibrium optical materials to achieve novel and hereto unexplored effects  \cite{Ghofraniha09,Conti10}, a {\it non-ergodic nonlinear optics} that is rooted in statistical mechanics, material science, and nonlinear wave propagation. Subwavelength propagation overcomes a basic limit to optical imaging and microscopy, i.e., that a light field can only propagate components of its spatial spectrum within the diffraction limit.  High frequency components that correspond to details comparable to and smaller than the wavelength normally form evanescent waves that simply do not propagate.  In stark contrast, in our predictions light leads to self-trapped beams of arbitrary intensity and widths, for which no diffraction limit holds.  These predictions are based on an intensity-independent nonlinearity, which can be interpreted as the consequence of a widely tunable refractive index accompanied by a transmission of evanescent fields.  Put differently,  a  nonlinearity-based meta-material \cite{Veselago68,EnghetaBook,Pendry00, Kivshar09, MunkBook}

\noindent {\it Intensity-independent PNR nonlinearity ---}
Our model system is a compositionally disordered and impurity-doped photorefractive relaxor ferroelectric (e.g., the KLTN \cite{Agranat92}).  When it is rapidly cooled below the characteristic Bruns temperature $T_B$, it exhibits polar nano-regions (PNR). These are highly polarizable randomly-distributed ferroelectric-like regions that form a dipolar glass and provide an enormous enhancement of the photorefractive nonlinear optical response but retain very limited scattering losses \cite{DelRe11}. The clue to the whole matter, originally discussed by Burns \cite{Burns83}, is that the optically-induced index of refraction change depends on the \textit{spatial average} of the \textit{square} of the crystal polarization $\mathbf{P}$. When the crystals display a zero polarization due to disorder (i.e., $\langle \mathbf{P} \rangle \approx 0$), averaging over disorder leads to a non-vanishing effect that depends exclusively on the mean-square of the polarization (i.e., $\langle |\mathbf{P}|^2 \rangle$).  Optical response is thus directly correlated to the underlying nature of the crystal \textit{fluctuations}, that become anomalously \textit{large} and dependent on the crystal \textit{history} in the out-of-equilibrium state. Specifically, averaging over the randomly oriented PNR \cite{Samara03, Bokov06}, the index of refraction perturbation is
\begin{equation}
\Delta  n_{PNR}=-\frac{n_0^3}{2} g   \epsilon_0^2 \chi_{PNR}^2 E_{DC}^2\text{,}
\label{dnf}
\end{equation}
where $n_0$ is the bulk refractive index of the isotropic (disordered) crystals,
$g$ is the relevant component of the second-order electrooptic tensor,
$\chi_{PNR}$ is the low-frequency electric response due to the PNR.
\indent The low-frequency electric field $E_{DC}$ is the space-charge field
expressed in terms of the optical intensity $I$ as \cite{Crosignani99, DelRe11}
$E_{DC}=-(K_B T/q)|\nabla I|/I$, with $T$ the crystal temperature, $K_B$ the Boltzmann constant and $q$ the elementary charge. 
Note that it is this dependence on $I$ that bestows on all effects their characteristic intensity independence. Non-ergodicity implies that $\chi_{PNR}$ will depend on the history  of the sample, so that the same crystal at the same temperature will display radically different nonlinear optical responses \cite{DelRe11}. Equation (\ref{dnf}) holds because  $\chi_{PNR}$ is several orders of magnitudes greater than the susceptibility of the crystals in the paraelectric phase $\chi_P$, so that  the nonlinear effect is mainly due to the PNR. Terms in the index perturbation that depend on the spontaneous polarization $P_0$ (i.e., in $\langle P_0^2 \rangle$) do not depend on the optical field, are negligible compared to the effects of the PNR, and are henceforth dropped.   
Finally, off-diagonal index modulation terms, that would amount to polarization rotation effects, are averaged out by the PNR disorder, and hence are negligible.  

\begin{figure}
\includegraphics[width=\columnwidth]{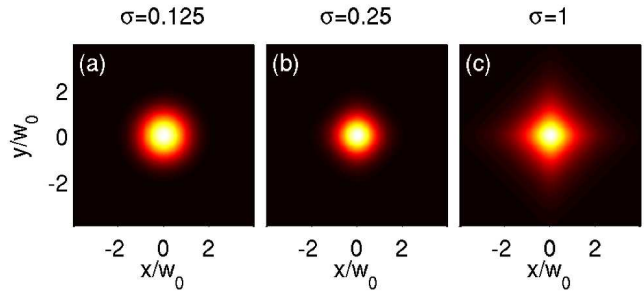}
\caption{ 
Scale free solutions for various values of $\sigma=(L^2/8\lambda^2)$.
We show the profile in (a) of the Gaussian solution ($\sigma=0.125$) and in (b,c) of the generalized solution Eq.(\ref{coshsol})
\label{figsquare}}
\end{figure}

\noindent {\it Scale-free self-trapped paraxial beams ---}
In the paraxial approximation, for a linearly polarized beam, the slowly-varying optical field $A$ ($|A|^2=I$ is the optical intensity)
obeys the nonlinear equation
\begin{equation}
2  i k \frac{\partial  A}{\partial z}  +\nabla^2_\perp  A-
\frac{L^2}{4\lambda^2} \frac{  (\partial_x I)^2+(\partial_y I)^2}{I^2} A=0\text{,}
\label{soliton_equation}
\end{equation}
where we have introduced the characteristic length 
\begin{equation}
L=4 \pi n_0^2  \epsilon_0 \sqrt{g} \chi_{PNR} (K_B T/q) \text{.}
\end{equation}
with $g>0$ (for commonly adopted ferroelectrics) and $k=\omega n_0/c$. The resulting model  admits analytical self-trapped solutions,  originally found in \cite{Crosignani99}
and experimentally investigated in \cite{DelRe11}
if condition $L\geq \lambda$
is  fulfilled.
For  $L= \lambda$ one has the exact Gaussian solution $A=a\exp(-i\beta z)$ with
\begin{equation}
a=A_0 \exp\left(-\frac{x^2+y^2}{w_0^2}\right),
\label{GaussianSolutions}
\end{equation}
with $\beta=2/k w_0^2$, the nonlinear correction to the wave-vector $k$. Remarkably, in Eq.(\ref{GaussianSolutions})  the waist  $w_0$ of the  soliton and  its amplitude $A_0$ are \textit{free independent parameters}.
When $L> \lambda$ a notable effect is that self-trapped  solutions
are given by $A=a e^{-i\gamma^2 \beta z}$ with
\begin{equation}
a=A_0 \left[ \cosh\left(\sqrt{2}\frac{x}{w_0}\right) \cosh\left(\sqrt{2}\frac{y}{w_0}\right)\right]^{-\gamma^2},
\label{coshsol}
\end{equation}
and $A_0$ and $w_0$ arbitrary constants (i.e., the ``existence curve''  is \textit{flat} \cite{DelRe09}), while  
\begin{equation}
\gamma=\sqrt{\frac{1}{\left(\frac{L}{\lambda}\right)^2-1}}\text{.}
\label{gamma}
\end{equation}
Interestingly,  as $L$  grows the beam {\it loses its radial symmetry},   developing a   square    like   profile.   In Fig.(\ref{figsquare}) we compare the two solutions.

\noindent {\it Suppression of evanescent waves and the absence of a diffraction limit ---}
We now consider scale-free solutions for beam waists comparable and \textit{smaller} than the wavelength.
Specifically, we use the Helmholtz equation, which generalizes the paraxial equation (\ref{soliton_equation}). The only approximation we implement with respect to the Maxwell equations is neglecting coherent vectorial coupling (e.g., the transfer of energy from a linearly polarized beam to the orthogonal polarization): this is expected to play a negligible role for  isotropic disordered crystals and for the  diffusive nonlinearity, which only depends on the local intensity and not on the beam polarization \cite{Crosignani99,DelRe09}.
The Helmholtz model reads as ($n=n_0+\Delta n$)
\begin{equation}
\nabla E+\left(\frac{\omega n}{c}\right)^2 E=0\text{,}
\label{Helmholtz}
\end{equation}
whose propagation invariant solution is written as $E=a\exp(i k_z z)$
where $k_z$ is the overall wave-vector in the $z$ direction
(different from its nonlinear perturbation $\beta$ in the paraxial model above). The scale-free Gaussian (\ref{GaussianSolutions}) is also solution 
of Eq. (\ref{Helmholtz}) with arbitrary amplitude $A_0$ and waist $w_0$, 
with 
\begin{equation}
k_z=\sqrt{\left(\frac{2\pi n_0}{\lambda}\right)^2-\frac{4}{w_0^2}}
\end{equation}
when the condition $L=\lambda$ is satisfied.
Note that this solution exists (i.e. $k_z$ is real) as long as
\begin{equation}
w_0> \frac{\lambda}{\pi n_0}
\end{equation}
that is, for a beam waist that is (within moltiplicative constants)  greater than the wavelength $\lambda/n_0$.

The key point is that more general solutions exist for $L>\lambda$, analogous to what occurs in the paraxial case.  These are still given by Eq.(\ref{coshsol}) where, however, the wavector is 
\begin{equation}
k_z=\sqrt{\left(\frac{2\pi n_0}{\lambda}\right)^2-\frac{4\gamma^2} {w_0^2}}\text{.}
\end{equation}
The corresponding lower limit for the waist is hence
\begin{equation}
w_0>\gamma\frac{\lambda}{\pi n_0}\text{.}
\end{equation}
The factor $\gamma$ plays a role similar to a Lorentz contraction
term in special relativity, even if with a different meaning. 
Specifically, for $L>\sqrt{2}\lambda$ the lower limit for the waist 
is scaled by a factor $\gamma<1$, and correspondingly beams with a width smaller than the wavelength can propagate in the medium without distortion.
Note that for this solution the evanescent waves are completely inhibited by the intensity independent scale-free nonlinearity,
which leads to the conclusion that arbitrarily-low power beams with size below the wavelength can propagate.

\noindent {\it Non-paraxial regime ---}
To investigate the formation of the two-dimensional (2D) soliton in the nonparaxial model
we consider the forward propagating projection of the Helmholtz equation, which is written as
\begin{equation}
i \partial_z E+k \sqrt{ 1+\frac{\nabla^2_\perp}{k_0^2}+2 \frac{\Delta n}{n_0}}E=0\text{.}
\label{forwHelm}
\end{equation}

Equation (\ref{forwHelm}) reduces, under suitable limits, to the well-known unidirectional propagation equations 
(see \cite{Kolesik02, Conti05PRL} and references therein) for the description of  nonlinear optics beyond the paraxial model. 
The basic difference here is that the nonlinear refractive index $\Delta n$ is retained under the square root, 
since our nonlinearity is intensity-independent and hence of the same order of the 
Laplacian even in the low intensity regions of the beam.  
After introducing the optical carrier with $E=A \exp(i k z)$ and the diffusive nonlinearity, the normalized dimensionless model reads as the non-paraxial normalized model
\begin{equation}
\begin{array}{l}
  i\partial_\zeta  \psi  +\frac{1}{\epsilon}\left[-1+\sqrt{1+\epsilon \nabla^2_{\xi,\eta} -2 \epsilon \sigma ({\bf v}\cdot{\bf v})}\right]\psi =0\text{,}\\
  {\bf v}|\psi|^2 +\nabla_{\xi\eta} |\psi|^2=0\text{,}
\end{array}
\label{soliton_equation_2}
\end{equation}
where we introduce the dimensionless variables $\xi=x/w_0$, $\eta=y/w_0$, and $\zeta=z/z_0$ 
with $z_0=k w_0^2$  the Rayleigh length. $\psi=A/A_0$ is the normalized optical field with $A_0$ an arbitrary constant, 
and ${\bf v}={\bf E}_{DC}/(K_B T/q)$ is the normalized space-charge field.
In Eq.(\ref{soliton_equation}) only two parameters appear, the degree of non-paraxiality $\epsilon$, which vanishes in the paraxial limit,
and the strength of the scale-free nonlinearity $\sigma$ ($\sigma=1/8$ as $L=\lambda$).

Equation (\ref{soliton_equation}) can be numerically solved by Taylor expanding the square root, giving
\begin{equation}
\partial_\zeta \psi=i\sum_{n=1}^{N} \binom{\frac{1}{2}}{n} \epsilon^{n-1} 
\left[\nabla^2_{\xi,\eta} -2 \sigma ({\bf v}\cdot{\bf v}) \right]^n \psi
\label{ordern}
\end{equation}
where $N$ is the order of approximation selected (for a fixed $\epsilon$) in order to a have a given precision.

\begin{figure}
\includegraphics[width=\columnwidth]{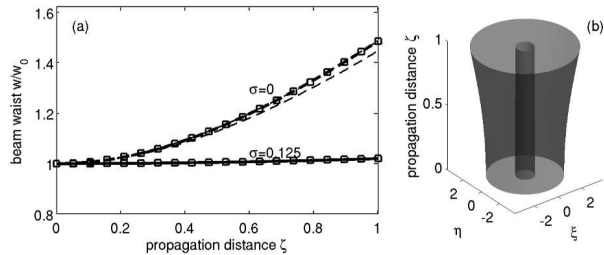}
\caption{ 
(a) Beam waist versus propagation for $\epsilon=0.05$ for
various orders of approximation of the nonparaxial equations (black line corresponds to standard paraxial models);
as $\sigma=0$ the nonparaxial terms lead to a more pronounced spreading;
for $\sigma=0.125$ an invariant propagation is attained at any non-paraxial order;
(b) three dimensional comparison between the case $\sigma=0$
and $\sigma=0.125$ for $N=3$.
\label{fig1}}
\end{figure}

To asses the existence of self-trapped scale-free solitons beyond the paraxial regime, we start by considering the linear 
diffraction regime ($\sigma=0$; $L<<\lambda$) in Fig.\ref{fig1}a, which shows the spreading of the waist of a Gaussian beam for an increasing order in the
the solution of Eq.(\ref{ordern}). For $\epsilon>0$ non-paraxial terms provide a more pronounced spreading if compared to the paraxial model (i.e., 
to $N=1$), and it is shown that for the case of $\epsilon=0.05$ corrections become inconsequential for $N>3$.

In Fig.\ref{fig1}b we show the propagation of the beam for $\sigma=1/8$ ($L=\lambda$), including higher order diffraction. In such a nonlinear case,
diffractionless propagation is achieved at any order and for any intensity.
Note that the resulting beam is propagation invariant at any order $N$ independently on the scale $w_0$ and on the intensity level,
thus showing the fact that the proposed scale free solutions are indeed stable and exist for ultra-thin beams.

\begin{figure}
\includegraphics[width=\columnwidth]{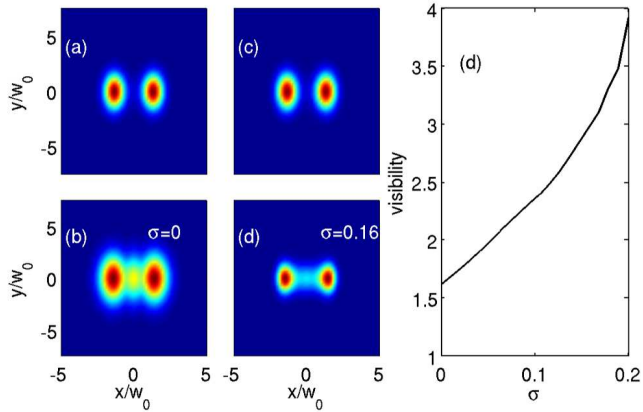}
\caption{ 
Intensity independent super-resolution:
(a,b) Propagation in the linear regime ($L=\sigma=0$) of a double spot (panel a) in the non-paraxial model Eq.(\ref{ordern}) at the third order ($N=3$) with $\epsilon=0.01$; (c,d) as in (a,b) with $\sigma=0.2$; (d) calculated visibility versus $\sigma$. The initial image in panels a,c is the same and included for the sake of comparison. \label{fig2}}
\end{figure}

\noindent {\it Enhanced visibility ---}
Next we numerically investigate conditions allowing super-resolution. We consider the
propagation of two parallel beams for $\epsilon=0.01$
(order $N=3$) for various values of the ratio $L/\lambda$. 
For a given input pattern we find that there exist an optimal value corresponding to mimimal input intensity distribution (i.e., image) distortion.
In Fig.\ref{fig2}, we consider the propagation of two parallel beams (shown in Fig.\ref{fig2}a,c)
for different values of $\sigma$. 
We compare the (Fig.\ref{fig2}b) linear propagation regime ($\sigma=L/\lambda=0$) and the case $\sigma=0.15$ (Fig.\ref{fig2}c); 
for $\sigma>0$ (corresponding to high cooling rates see \cite{DelRe11}) the visibility radically increases. 
In Fig.\ref{fig2}d we show the frange visibility versus $\sigma$, 
calculated as the ratio between the intensity peak value and the value at the center of the beam:
the contrast increases (ideally diverges) when the visibility is higher.
This result shows that even for beam sizes in the non-paraxial regime, at sufficiently high cooling rate, it is possible to propagate images without loss of information.
We stress that these dynamics are attained at any intensity level,
and hence represent a completely new regime in optical propagation,
where the laws of diffraction and interference are largely modified;
the strength of nonlinearity being determined by $\sigma$.
For $L>>\lambda$ ($\sigma>>1/8$), a Gaussian beam experiences a strong intensity-independent self-focusing, 
even below the non-paraxial regime (not reported).

\noindent {\it Three-dimensional subwavelength localization ---} A notable property is the existence of three-dimensional localized light bullets.
Specifically the three-dimensional (3D) Helhmholtz equation for the diffusive nonlinearity can be cast as a ``nonlinear'' eigenvalue problem:
\begin{equation}
-\frac{\nabla^2 E}{E}+\left(\frac{L}{\lambda}\right)^2 \left( \frac{\nabla |E|^2}{2|E|^2}\right)^2 =k^2 
\label{h3d}
\end{equation}
Eq. (\ref{h3d}) admits an {\it exact} 3D Gaussian solution $A=A_0 \exp\left[-\left(x^2+y^2+z^2\right)/w_0^2\right]$ when $L=\lambda$, for any $A_0$ and
when $w_0=\sqrt{3/2\pi} \lambda/n_0$. In distinction to previous results, these solutions are not spatially scale-free and may only have a fixed waist, comparable with the wavelength in the medium.
For $L>\lambda$ a solution exists (more general ones exist and will be reported elsewhere) and is given by
\begin{equation}
A=A_0 \left[ \cosh(\sqrt{2} x/w_0) \cosh(\sqrt{2} y/w_0) \cosh(\sqrt{2} z/w_0) \right]^{-\gamma^2}
\end{equation}
with $\gamma$ as above (Eq.(\ref{gamma})) and the waist $w_0=\gamma \sqrt{3/2 \pi} \lambda/2\pi n_0$.
These 3D localized solutions have a waist smaller that the wavelength when $\gamma<1$ ($L>\sqrt{2}\lambda$).
They represent a novel form of light localization at any intensity level and with size
comparable or smaller than the wavelength. We are not aware of other known light localizations, which can be described by an exact solution; it is a novel kind of bound state between the photo-induced charges and light which may be used to store information.

\noindent {\it Metamaterials of nonlinear origin ---}
In standard optics, a Gaussian beam with waist $w_0$ has a spectral bandwidth of the order of $1/w_0$, and the minimum waist 
such that the spectrum is contained in the Ewald circle (i.e., without evanescent waves) is given by $\lambda/n_0$;
in the scale-free regime here considered, due to the non-ergodic phase of glassy ferroelectrics, such a minimum waist is given by $\gamma\lambda/n_0$.
The medium hence exhibits (for $L>>\lambda$) an effective refractive index
\begin{equation}
n_{eff}=\frac{n_0}{\gamma}=n_0 \sqrt{\left(\frac{L}{\lambda}\right)^2-1}\cong n_0 \frac{L}{\lambda}>>n_0\text{.}
\end{equation}
The PNR effect is therefore equivalent to an (intensity {\it independent}) refractive index,
which can be largely tuned and increased, such that beams propagate without evanescent waves
and without relevant scattering and absorption losses.
This shows that the specific nonlinearity here considered is able to provide those features
that are the building blocks for the modern research on meta-materials, from a completely different perspective.

We acknowledge support from the CINECA-ISCRA parallel computing initiative.
The research leading to these results has received funding from the
European Research Council under the European Community's Seventh Framework Program 
(FP7/2007-2013)/ERC grant agreement n.201766, and from the Italian Ministry of Research (MIUR) through the "Futuro in Ricerca" FIRB-grant PHOCOS - RBFR08E7VA. Partial funding was received
through the SMARTCONFOCAL project of the Regione Lazio. 


\end{document}